\documentclass[%
 aip,
 amsmath,amssymb,
 reprint,%
]{revtex4-1}

\usepackage{graphicx}%
\usepackage{dcolumn}%
\usepackage{bm}%
\usepackage[utf8]{inputenc}
\usepackage[T1]{fontenc}
\usepackage{mathptmx}

\begin{document}

\preprint{}

\title[Alignment-free cryogenic optical coupling to an optomechanical crystal]{Alignment-free cryogenic optical coupling to an optomechanical crystal}

\author{Timothy P. McKenna}
  \altaffiliation{These authors contributed equally.}
\author{Rishi N. Patel}%
  \altaffiliation{These authors contributed equally.}
  \author{Jeremy D. Witmer}%
  \altaffiliation{These authors contributed equally.}
\author{Rapha\"el Van Laer}%
  \altaffiliation{These authors contributed equally.}
 \author{Joseph A. Valery}%
\author{Amir H. Safavi-Naeini}
  \email{safavi@stanford.edu.}
\affiliation{ 
Ginzton Laboratory, Stanford University, 348 Via Pueblo Mall, Stanford, California 94305, USA
}%

\date{\today}%

\begin{abstract}
The need for highly accurate, labor-intensive optical alignment has been a major hurdle in our ability to leverage the power of complex photonic integrated circuits. There is a strong need for tolerant and passive alignment methods that enable interrogation at any point in a photonic circuit. Various promising and scalable photonic packaging techniques have been under development, but few methods compatible with low-temperature operation have been reported. Here, we demonstrate alignment-free $25\%$ coupling efficiency from an optical fiber to a silicon optomechanical crystal at 7 mK in a dilution refrigerator. Our coupling scheme uses angle-polished fibers glued to the surface of the chip. The technique paves the way for scalable integration of optical technologies at low temperatures, circumventing the need for optical alignment in a highly constrained cryogenic environment. The technique is broadly applicable to studies of low-temperature optical physics and to emerging quantum photonic technologies.
\end{abstract}

\maketitle

\section{\label{sec:level1} Introduction}

The rise of integrated optical systems has motivated the development of numerous approaches to efficiently couple light from a fiber's  $\approx 10 \, \mu \text{m}$ diameter optical mode into sub-micron confined channels on the surface of a chip\cite{Vermeulen2018}.  The emerging application space of quantum sensors, computers, and communication systems necessitates its own new classes of integrated optical devices for quantum communications and control. A key challenge is to efficiently couple light in and out of these devices. Since quantum systems often have components that operate at extremely low temperatures, optical systems operating at these temperatures require coupling schemes that are both scalable and immune to the thermal stresses induced by thermal cycling. In this work, we demonstrate an approach to coupling light in and out of a silicon photonic chip at millikelvin temperatures that is robust, requires no in-situ alignment and is compatible with numerous quantum technologies\cite{Schoelkopf2008,Sipahigil2016,Gould2016,Schroder2016,Witmer2017}.

Common methods to address the optical input/output barrier include inverted edge tapers, evanescent couplers, and grating couplers\cite{Vermeulen2018,Benedikovic2014a,Kopp2011}. Tapered fiber coupling, where a tapered fiber evanescently couples to an on-chip waveguide, has been used to achieve broadband efficient coupling at both room and cryogenic temperatures\cite{Srinivasan2007,Groblacher2013a,Macdonald2015a}. The technique uses specially formed fiber tapers and requires in-situ alignment at cryogenic temperatures, which limits the number of input-output channels and drastically increases the cost and complexity of the apparatus. Moreover, at low temperatures and in vacuum, the technique is subject to vibrational noise and power-dependent instabilities. Edge coupling has been demonstrated at cryogenic temperatures \cite{Patel2018,Meenehan2014,Riedinger2018} but also requires in-situ alignment.

 Another common coupling technique utilizes gratings patterned on the surface of a chip. Well-designed grating couplers can in principle convert over $90\%$ of incoming light into a waveguide on the chip over a bandwidth of several THz with reflections below $1\%$ and using a footprint of merely tens of square microns\cite{Taillaert2006,Benedikovic2014a,Vermeulen2018,Notaros2016,Son2018,Michaels2018}. They allow for optical access to any point on a wafer while requiring only micron-level alignment accuracy, drastically reducing the resources necessary to interrogate complex optical circuitry. Moreover the relatively high tolerance to misalignment makes grating couplers a promising candidate for cryogenic coupling where thermal stresses increase the likelihood of the fiber moving with respect to the grating coupler.  As an example of this, Shainline \textit{et al.}\cite{Shainline2017} recently demonstrated a technique in which an optical fiber can be aligned to a grating coupler with the assistance of an SU-8 collar, and achieved coupling efficiencies of 21\% at a temperature of $1~\text{K}$ without requiring low-temperature re-alignment.%

Light emitted from a grating coupler is often mode-matched to SMF-28 fiber so that a cleaved fiber facet can be used to efficiently collect the radiation. The fiber, or fiber assembly, must be held at design-determined angle to the normal of the chip surface for efficient coupling. In the work by Shainline \textit{et al.}\cite{Shainline2017}, this was accomplished by extra layers of processing to build a supporting structure on the surface for the fiber, and a larger assembly around the chip to hold the fiber at an angle. The extra processing steps make that approach difficult to combine with released optomechanical structures. An alternative approach is to use angle-polished fibers\cite{Snyder2013a,Li2014} that can rest horizontally on the chip surface, obviating the need to support a fiber at an angle and leaving space for electrical connections. Total internal reflection occurring at the fiber/air interface sends light into an on-chip grating coupler at the correct angle for efficient coupling (see Figure ~\ref{fig:device}a). 

Here, we propose and develop a fiber-chip coupling technique, based on grating couplers and angle-polished fibers, and use it to demonstrate cryogenic measurements down to millikelvin temperatures. The technique is scalable in the sense that it provides access to many optical input and output ports on a chip at cryogenic temperatures without requiring a corresponding number of stages and control wires for preserving alignment. We demonstrate the technique by measuring a silicon optomechanical crystal in a dilution refrigerator at $7 \, \text{mK}$. The technique has good yield and the packaged devices remain stable over many cooling cycles. It circumvents the need for time-consuming manual alignment in a constrained cryogenic environment and paves the way for complex circuits with many inputs and outputs operating at low temperatures for integrated quantum optical \cite{Silverstone2016,Rudolph2017}, optomechanical \cite{VanLaer2015,Kittlaus2015,Patel2017,Patel2018,Safavi-Naeini2019}, and electro-optic \cite{Bochmann2013,Witmer2017,VanLaer2018} devices.

\section{\label{sec:level1} Procedure}

\subsection{\label{sec:level2}Device Fabrication}

\begin{figure}[h]
\centering
\includegraphics[scale=0.9]{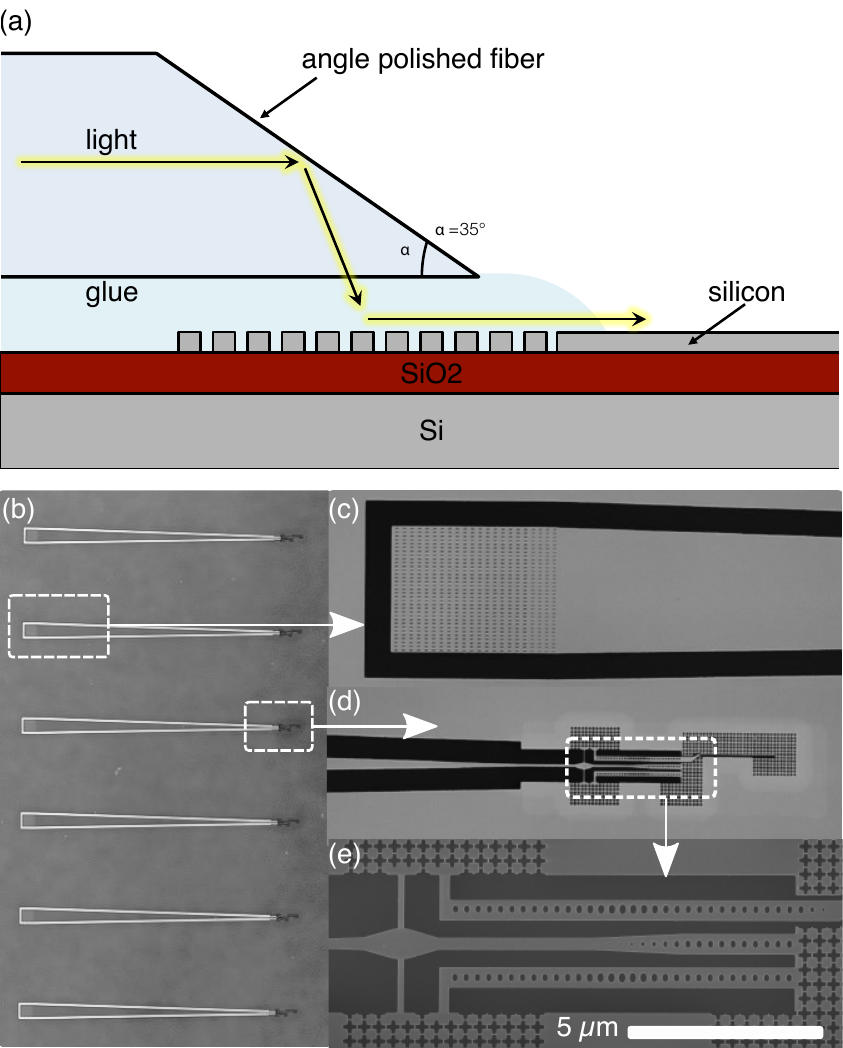}
\caption{(a) Coupling scheme using angle cleaved fiber and grating coupler fabricated on an SOI platform. (b) Laser confocal microscope image showing array of devices on a SOI chip. Each device consists of a grating couplers attached to a suspended optomechanical crystal. (c) Top view of the grating coupler. (d) Suspended optomechanical crystal devices. (e) SEM of optomechanical crystals positioned near a central waveguide for evanescent optical coupling.}
\label{fig:device}
\end{figure}

We fabricate gratings couplers attached to optomechanical crystals using a two-step silicon-on-insulator (SOI) process. First, we pattern fine features in silicon by ebeam lithography and a Cl$_{2}$-based etch \cite{Patel2017,Patel2018}. Second, we selectively suspend the optomechanical crystals but not the grating couplers by optical lithography and a buffered HF etch \cite{VanLaer2015,VL2016}. We fabricate an array of such devices with a parameter sweep in the overall scaling. This allows us to frequency match the optical cavity resonance with the grating coupler response. An image of a device array is shown in Figure~\ref{fig:device}, along with an image of the fiber-coupled device.

The grating coupler consists of a square array of rectangular holes fully etched into the 220 nm silicon slab according to a metagrating design \cite{Benedikovic2014a} with simulated efficiencies up to $60\%$. It has an impedance-matching set of smaller holes to reduce reflections off the grating (Figure 1(c)).

\subsection{\label{sec:level2}Fiber Gluing Procedure}

To couple light into the on-chip grating couplers, we use angle-polished SMF-28 fibers (Chuxing Optical Fiber Application Technologies Ltd.) with a polish angle of approximately $35^\circ$.  The fiber is positioned horizontally above the surface of the chip.  The light in the fiber experiences total internal reflection at the fiber facet and is reflected down through the bottom of the fiber, impinging on the chip surface with an angle of approximately $27^\circ$ from normal when no glue is present. This angle was chosen to recycle light reflected off the handle silicon \cite{Benedikovic2014a}. The coupling scheme is illustrated in Figure \ref{fig:device}(a).

\begin{figure*}[htbp]
  \centering
  \includegraphics[scale=0.95]{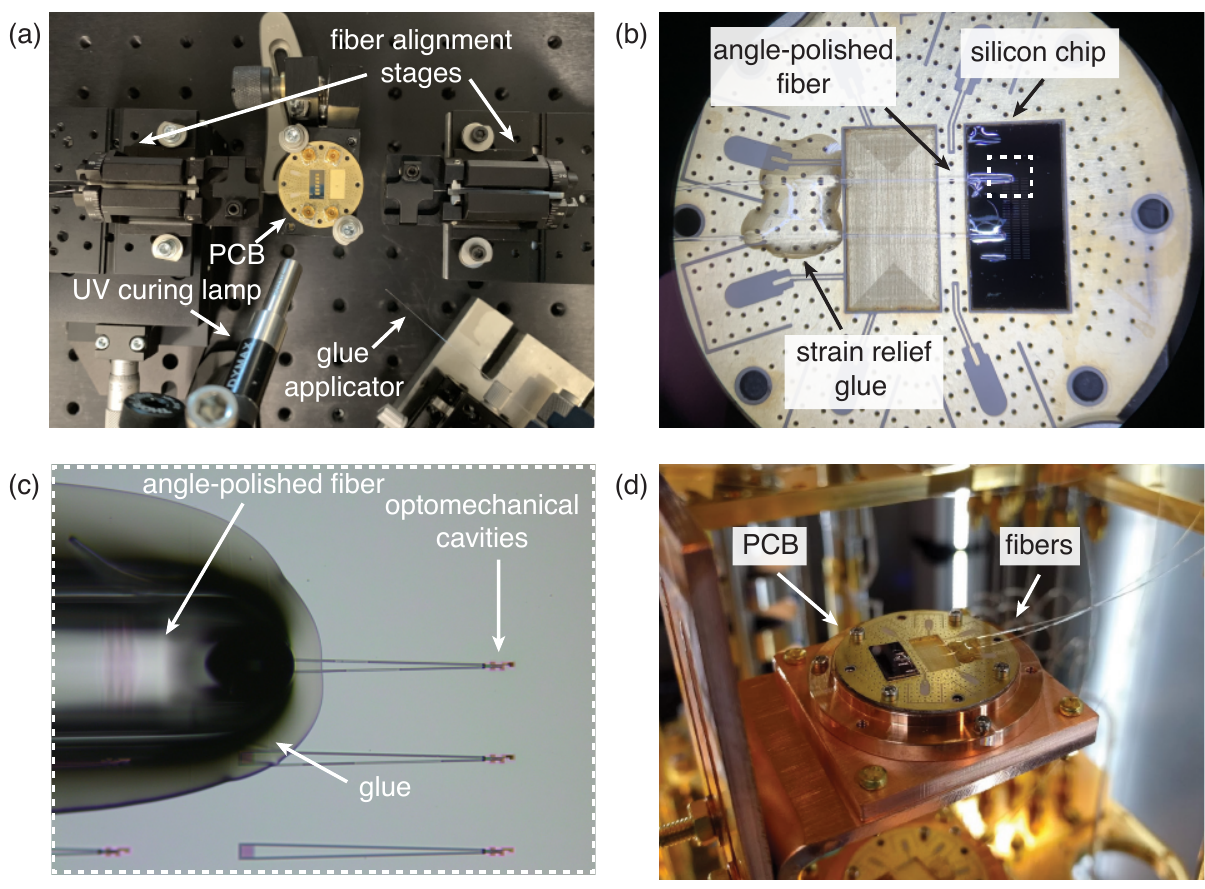}
\caption{(a) The fiber gluing setup.  The PCB with the photonic chip is fastened to a mirror mount.  Micrometer stages with adjustable fiber clamps are used to control the fiber X, Y, Z, roll and yaw. A spare optical fiber on a separate XYZ stage stack is used to apply the glue. (b) The PCB viewed under a microscope. (c) A zoomed-in image of the dashed region in part (b).  The angle-polished fiber is aligned to the grating coupler of the optomechanical device.  The cured glue ``puddle'' surrounding the fiber is visible.  (d) The PCB mounted at the mixing chamber of the dilution refrigerator.}
\label{fig:gluing_procedure}
\end{figure*}

To prepare for fiber gluing, we first fasten the chip to a custom-made copper printed circuit board (PCB) using a GE Varnish/ethanol mixture.  The PCB is then mounted in our fiber gluing setup, shown in Figure \ref{fig:gluing_procedure}(a).

We align the angle-polished fiber to the device of interest before applying the glue.  Next, we adjust the fiber X, Y, Z, yaw and roll in order to optimize the optical reflection signal from the grating coupler. We have also used this procedure for transmission devices with multiple couplers.  We then apply a small drop of glue (Norland Optical Adhesive 88) to the fiber, about 0.5 to 1 mm away from the tip of the fiber.  To do this we use a spare piece of SMF-28 fiber as an applicator and control the applicator position with an XYZ micrometer stage stack. After touching down the glue-tipped applicator, we move the applicator until the glue wicks underneath the fiber and covers the grating coupler by capillary action.

After the glue is applied, we find it necessary to adjust the fiber X and Y position to reoptimize the reflected signal.  After reoptimizing the position, we cure the glue using a UV curing spot lamp (DYMAX BlueWave 75).  We use the lamp at full power and do 6 cures of 30 seconds each, at a distance of about 3 cm and from a few different angles. 

Once the fiber is glued to the chip, we apply a second, much larger drop of glue to fasten the fiber to the PCB and provide strain relief.  We make sure that this second drop covers a portion of the fiber where the fiber coating is still intact to provide strength.  We cure this glue in the same way as the first drop.  After curing the glue, we typically leave the chip in an air environment for about 12 hours, allowing the glue to age and achieve better adhesion before installing it in the dilution refrigerator.

Once the fiber gluing process is complete, the PCB holding the chip is installed at the mixing chamber of a Bluefors dilution refrigerator (see Figure \ref{fig:gluing_procedure}(d)).  The dilution fridge is fitted with a series of optical fibers that provide optical access from room temperature down to the mixing chamber. 

\section{Experimental Results}

\subsection{Predicting Spectral Shifts}
Our optical devices experience two spectral shifts during the gluing process and subsequent cooldown. First, once the glue is applied, the grating coupler spectrum undergoes a large redshift of about 91 nm (not shown). This shift stems from the refractive index of the glue ($\approx$ 1.56) causing an increase in the Bloch mode index of the grating coupler. Experiments performed with oxide-clad grating couplers do not exhibit this redshift. Since the glue stays localized near the coupler, the optomechanical crystal does not undergo a frequency shift upon application of the glue.  Second, both the grating coupler spectrum and the optical resonance frequency of the optomechanical crystal experience a blueshift as the chip is cooled from $300~\text{K}$ to millikelvin temperatures (Figure \ref{fig:exp_results}).  

To account for these shifts, we initially measure the optical reflection spectra in air before applying any glue. We use these spectra to predict device behaviour after gluing and cooling to low temperatures. Devices of interest are selected for gluing based on their predicted behavior after the cooldown.

To understand and predict the spectral shifts quantitatively, we develop a few analytical and finite-element models. For the grating coupler, the phase-matching condition between light incoming at angle $\theta=27^{\circ}$ and a guided wave with wavevector $\beta$ is
\begin{equation}
\beta(\omega)=k_{0}(\omega)n_{t}\sin{(\theta)}   + G  
\end{equation}
with $\beta = k_{0}n_{\text{eff}}$ the Bloch wavevector of the guided optical mode, $k_{0}=\omega/c$ the free-space wavevector, $n_{\text{eff}}\approx 2.4$ the effective Bloch index, $n_{t}$ the refractive index of the medium atop the grating, which is either air ($n_{t}=1$) or glue ($n_{t}=1.56$) and the grating wavevector $G=2\pi/\Lambda$ with $\Lambda=0.81 \, \mu\text{m}$ the grating pitch. The phase-matching condition rewritten in terms of the center wavelength $\lambda_{c}$ of the grating is
\begin{equation}
    \lambda_{c} = \Lambda(n_{\text{eff}}(\lambda_{c}) - n_{t}\sin{(\theta)})
\end{equation}
We are interested in perturbations of the center wavelength given perturbations in (1) the refractive index $n_{t}$ of the medium atop the grating and (2) the refractive index $n_{\text{Si}}$ of the silicon. We must take into account the dispersion $n_{\text{eff}}(\lambda_{c})$ of the Bloch mode's effective index to predict the size of such perturbations. Taylor-expanding the dispersion and truncating to first-order yields
\begin{equation}
    \delta \lambda_{c} = \delta n_{\text{eff}} \frac{\Lambda}{1 + \frac{\Lambda}{\lambda_{c}}(n_{g} - n_{\text{eff}})}
\end{equation}
with $n_{g} = n_{\text{eff}} - \lambda \frac{\partial n_{\text{eff}}}{\partial \lambda} \approx 3.5$ the group index of the Bloch mode. This expression assumes a fixed $n_{t}\sin{\theta}$, which holds in all cases of interest here since horizontal momentum is conserved at interfaces between the fiber and the glue or air.

First, we apply equation (3) to consider the shift of the grating coupler center wavelength as we apply the glue. Our finite-element models predict a shift in the Bloch index of $\delta n_{\text{eff}}\approx 0.21$. This yields a predicted redshift $\delta \lambda_{c}\approx 102 \, \text{nm}$ in reasonable agreement with the observed redshift $\delta \lambda_{c}\approx 91 \, \text{nm}$.

Second, as we cool down from $300 \, \text{K}$ to $7 \, \text{mK}$, the refractive index of silicon drops\cite{Frey2006a} from $3.486$ to $3.453$ so $\delta n_{\text{Si}} = -0.033$. This perturbs the Bloch index as $\delta n_{\text{eff}} \approx -0.025$, leading to predicted blueshift $\delta \lambda_{c} \approx -12.7 \, \text{nm}$ agreeing reasonably with the observed shift of the grating coupler, $\delta \lambda_{c} \approx -10.5 \, \text{nm}$ (Figure 3). We neglect the temperature-dependence of the refractive index of the glue and the silicon dioxide in this calculation.

Third, we can model the optomechanical crystal as a cavity made of a waveguide supporting an optical mode with group index $n_{g} \approx 4$.  The shift for such a cavity is given by
\begin{equation}
    \delta \lambda_{c} = \lambda_{c} \frac{\delta n_{\text{eff}}}{n_{g}},
\end{equation}
giving a predicted blueshift of $\delta \lambda_{c} \approx -13 \, \text{nm}$. Finite-element modeling of the optomechanical crystal cavity optical mode confirms this: treating the refractive index shift as a perturbation\cite{Chan2011,Frey2006a} yields $\delta \lambda_{c} = -12.6 \, \text{nm}$ in a finite-element simulation -- in close agreement with the observed blueshift (Figure 3).

\subsection{Optical Measurements}

After gluing, the devices are mounted inside the cryostat where characterization is performed at different temperatures. We send an optical pump from a laser tuned to approximately 1550 nm to the glued angle-polished fiber. Since the device input coupling waveguide is terminated in a photonic crystal mirror whose bandwidth exceeds the grating coupler bandwidth, and whose reflection coefficient can be taken to be unity, knowledge of the external system efficiency allows us to calculate the single-pass efficiency of the grating coupler.  

\begin{figure}[h]
\centering
\includegraphics[scale=0.9]{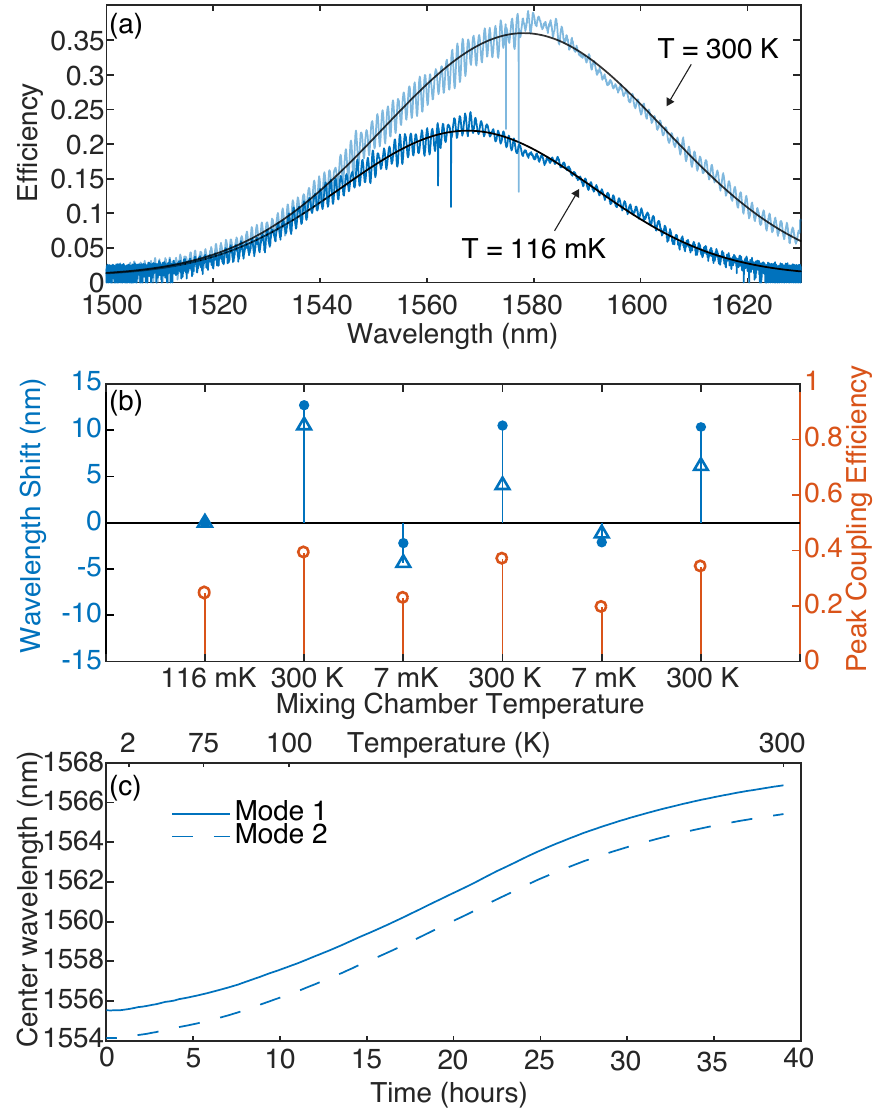}
\caption{(a) The light (dark) blue shaded curves show the single pass device efficiency at room (cryogenic) temperature. The black lines are Gaussian fits. (b) Mode wavelength shift and device coupling efficiency across cooldowns, referenced to the first dataset at $T = 116~\text{mK}$. The solid circles correspond to photonic crystal cavity mode shifts, while the open triangles denote the grating coupler shift. The open red circles show the peak coupling efficiency. (c) Cavity mode center frequency shift versus time over the course of a warmup. On the top axis we show the temperature of the mixing chamber stage.}
\label{fig:exp_results}
\end{figure}

We summarize the observed device efficiencies over three thermal cycles in Table~\ref{Table:efficiencies}. We show the reflection spectra for one device at room temperature and at $100~\text{mK}$ in Figure~\ref{fig:exp_results}(a). The broad response in the reflection spectrum is due to the grating coupler, while the narrow dips are due to the optical cavity resonances. There are two such resonance per device as there are two optomechanical crystals present per coupling waveguide (Figure 1(e)). In our devices we observe optical quality factors on the order of $10^5$, with no significant variation across cooldowns. An overall blueshift of the cavity wavelength and grating spectrum of approximately $12.6~\text{nm}$ is observed, in excellent agreement with finite-element simulations of our structure (see above).

Occasionally a glued fiber comes loose during a cooldown and the optical reflection signal is lost outright. We believe thermal contraction of the glue is responsible for these occasional failures. The overall rate of such failures is < 25\%, and if a given fiber survives its first cooldown, we find that it is highly likely to survive subsequent cooldowns. In this study device 1 failed after one thermal cycle as shown on the first row of Table~\ref{Table:efficiencies}. However, all other devices survive multiple cycles. In particular, device 3 remained intact for 10 cooldowns over the course of 6 months before being removed from the dilution fridge, without suffering substantial degradation in its optical coupling efficiency.  

\begin{figure}[h]
\centering
\includegraphics[scale=0.9]{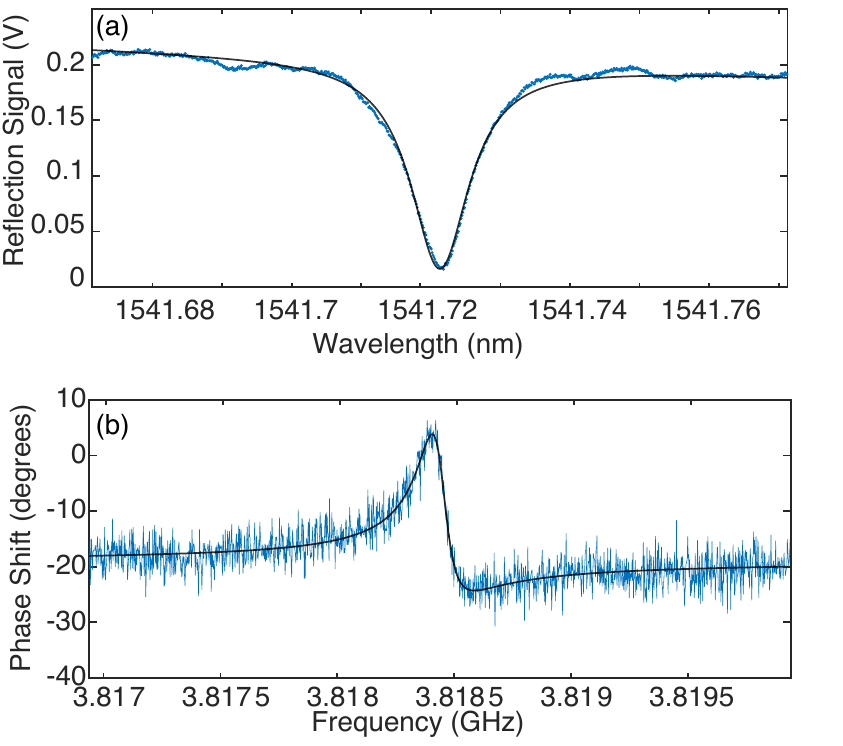}
\caption{(a) Optical mode reflection data of a glued device, with $Q\approx1.5\cdot10^5$. (b) Phase shift of a weak optical probe tone incident on the optomechanical crystal. We extract the optomechanical parameters of the system (see main text) from this dataset.}
\label{fig:exp_results_mechanics}
\end{figure}

We make optomechanical measurements of the devices under continuous wave laser driving. For a device cooled to $T=100$ mK, we monitor the 3.8 GHz mechanical mode and measure a mechanical quality factor $Q \approx 2\cdot 10^{5}$ and a single photon optomechanical coupling rate $g_{0}/2\pi = 726$ kHz (Figure \ref{fig:exp_results_mechanics}). For an input laser power $P\approx8 \mu\textrm{W}$ we reach a cooperativity of $C\approx0.8$ -- sufficient for itinerant state-transfer applications. These parameters are comparable to those used in recent heralded quantum entanglement experiments with mechanical systems\cite{Riedinger2018}. 

Further improvements to our devices should allow detection efficiencies of $60\%$ and higher to be attained\cite{Benedikovic2014a,Notaros2016,Vermeulen2018,Son2018,Michaels2018}, approaching typical efficiencies in experiments using lensed fibers and adiabatic edge couplers~\cite{Patel2018}. One promising application of our packaging technique is the ability to simultaneously monitor multiple devices. In such schemes, the number of devices that can be probed is only limited by chip space, and the number of fiber feedthroughs on the dilution fridge. This is in contrast to using lensed fibers mounted on stages, where independent control over multiple fibers quickly becomes infeasible. Finally, it is important to note that the grating coupler must be frequency matched to the cavities of interest in any given experiment. However, this problem is readily solved by the fabrication of arrays of devices, and rapid large area testing at room temperature. Using the relations shown in this paper for wavelength shifts due to gluing and thermo-optic effects, we can deterministically predict the cavity and grating wavelength post-cooldown and select suitable devices for gluing.

\begin{table}
\centering
\caption{Single-pass optical coupling efficiency for four devices over three thermal cycles. Device 1 failed after the first thermal cycle, although we find all other devices remain intact for multiple cycles.}
\label{Table:efficiencies}
\begin{tabular}{cccccccc}
Device & T = 300 K & 116 mK & 300 K & 7 mK & 300 K & 
7 mK & 300 K \\
\hline
$\eta_{1}$ & 0.29 & 0.21 & 0.26 & 0.00 & 0.00 & 0.00 & 0.00 \\
$\eta_{2}$& 0.36 & 0.25 & 0.40 & 0.23 & 0.37 & 0.20 & 0.34 \\
$\eta_{3}$ & 0.28 & 0.20 & 0.22 & 0.21 & 0.22 & 0.16 & 0.22 \\
$\eta_{4}$ & 0.25 & 0.20 & 0.24 & 0.19 & 0.24 & 0.21 & 0.25 \\
\end{tabular}
\end{table}

\section{Conclusion}
We demonstrate alignment-free coupling with $25\%$ coupling efficiency from an optical fiber to a silicon optomechanical crystal at 7 mK in a dilution refrigerator. The developed angle-polished fiber gluing technique has good yield, and the packaged devices remain stable over many cooling cycles. In the future, the technique can be expanded to individually monitor many devices on a single chip in a cryogenic environment. Additionally, the gluing principle is not limited to the use of grating couplers and can be extended to other coupling schemes such as tapered fiber and edge coupling. 

\begin{acknowledgments}
This work was funded by the ARO/LPS CQTS program and NSF ECCS-1708734. Part of this work was performed at the Stanford Nano Shared Facilities (SNSF) and Stanford Nanofabrication Facility (SNF) which are supported by the National Science Foundation under award ECCS-1542152. A.S.N. acknowledges the support of a David and Lucile Packard Fellowship. R.V.L. acknowledges funding from VOCATIO and from the European Union's Horizon 2020 research and innovation program under Marie Sk\l{}odowska-Curie grant agreement No. 665501 with the research foundation Flanders (FWO). J.D.W. acknowledges support from a Stanford Graduate Fellowship. R.N.P. is supported by a National Science Foundation Graduate Research Fellowship under grant no. DGE1656518. R.V.L. thanks Roel Baets and Dries Van Thourhout for helpful discussions.
\end{acknowledgments}

%
%

\end{document}